\documentclass[alpha-refs]{wiley-article}
\usepackage{siunitx}
\usepackage{gensymb}
\usepackage{multirow}
\usepackage{graphicx}
\papertype{Original Article}
\paperfield{Journal Section}

\title{Similarity indexing \& GIS analysis of air pollution}

\abbrevs{DSI, Delhi Similarity Index.}

\author[1\authfn{1}]{Purusharth Saxena}
\author[2\authfn{1}]{Madhu Kashyap Jagadeesh}

\contrib[\authfn{1}]{Equally contributing authors.}

\affil[1]{Department, Institution, City, State or Province, Postal Code, Country}
\affil[2]{Department of Physics,  Jyoti Nivas College,
  Bengaluru, Karnataka, 560029, India}

\corraddress{Madhu Kashyap Jagadeesh, Department of Physics,  Jyoti Nivas College, Bengaluru, Karnataka, 560029, India}
\corremail{kas7890.astro@gmail.com}

\fundinginfo{None}

\runningauthor{Purusharth and Kashyap}

\begin{document}

\maketitle

\begin{abstract}
Pollution has become a major threat in almost all metropolitan cities around the world. Currently atmospheric scientists are working on various models that could help us understand air pollution. In this paper we have formulated a new metric tool called Delhi Similarity Index (DSI). The DSI is defined as the geometrical mean of the trace gases such as: ozone, sulfur-dioxide and carbon-monoxide, which ranges from 0 (dissimilar to Delhi) to 0.9-1 (similar to Delhi). The limitation of the tool concerning the result of the nitrous-di-oxide data set is also analyzed. Also, the GIS projections of PM 2.5 role for Indian cities are graphically represented. The DSI results from 2011 to 2014 data show that Bengaluru is in the threshold of becoming as polluted like Delhi with values varying from 0.8 to 0.9 (i.e. 80-90\%) and Jungfraujoch with a 0.65 to 0.7 (i.e. 65-70\%).

\end{abstract}

\section{Introduction}
The pollution in metropolitan cities is leading towards numerous modeling and analysis studies. The research about air pollution, and controlling techniques are suggested to avoid the rise of pollution [\cite{Ratti} and \cite{Beelen}]. But New Delhi the capital city of India is infamous for its pollution, which has risen to such an extent that it is posing a non-negligible health hazard for the permanent residents of the city [\cite{Chhabra}, \cite{Cropper}, \cite{Agarwal}, \cite{Chhabra}, \cite{Cropper} \& \cite{Kumar}].
Despite constant research and concerns to government, the air quality in New Delhi is degrading and has reached a critical point [\cite{Kumar} and \cite{Agarwal}]. According to the Air Quality Index (AQI), the urban areas are facing major threat, consequently the trend study and analysis of AQI is going on at various places (Ex: \cite{Lan}).
In this research we use the evolved similarity index formula of \textit{Bray-Crutis} of 1957, which is applied in various studies such as ecology (\cite{Orloci}), marine (\cite{Schulz}), and astrophysics (\cite{Schulze-Makuch} and \cite{Madhu}). A new indexing measure to check the similarity of air pollution in different cities (such as: Bengaluru, New Delhi and Jungfraujoch) with respect to the New Delhi air pollution is introduced, which is called Delhi Similarity Index (DSI). 
DSI is basically defined as the geometrical mean of the trace gases such as: ozone, sulfur-dioxide and carbon-monoxide, which ranges from 0 (dissimilar to Delhi) to 0.9-1 (similar to Delhi). For past several years, Delhi's PM 2.5 (Particulate Matter $\leq 2.5$ microns) level has crossed the threshold of WHO standards atleast a few times in an year, as studied by \cite{Tiw}. Various studies have been performed over the atmospheric and surface pollution of Delhi: \cite{Ghude-Jain}, \cite{Srivastava}. Geographic Information System (GIS) has also been employed to study the spread and compare the pollution over time within a city. \cite{Matejicek}, \cite{Jerrett}. Hence in this paper we have dedicated a section to understand PM 2.5 trends in New Delhi and Bengaluru along DSI analysis.\\
The paper structure is as follows: Section 2 contains the formulation of DSI, Section 3 is about data collection along with the geo-political description of stations, section 4 deals with the Results of DSI, section 5 has the PM 2.5 analysis of Bengaluru and New Delhi, and section 6 is concluded by the discussion and conclusion.

\section{Formulation of Delhi Similarity Index (DSI)}

The Bary-Curtis' similarity index from 1957 is used in this formulation with the indices division of  0.2 intervals \cite{Bloom}, the classifications are as follows: very low, low, moderate, high and very high similarity regions. 
Here, we define the threshold to very high similarity region with the  threshold of $V=0.8$. Defining the physical limits $x_a$ and $x_b$ of the permissible variation of a variable with respect to $x_0$ (i.e. $x_a<x_0<x_b$), we calculate the weight exponents for the lower $w_a$ and upper $w_b$ limits,
\begin{equation}
w_a = \frac
{\ln{V}}
{\ln\left[1-\left|\frac{x_0-x_a}
{x_0+x_a}\right|\right]}
\,,\quad 
w_b = \frac
{\ln{V}}
{\ln\left[1-\left|\frac{x_b-x_0}{x_b+x_0}\right|\right]}
\,,\quad 
\label{eq:weight_exponent}
\end{equation}
The average weight is found by taking the geometric mean of $w_a$ and $w_b$,
 \begin{equation}
w_x=\sqrt{{w_a}\times {w_b}}\,.
\label{eq:geom_mean}
\end{equation}

The Delhi Similarity Index is defined, as (for the entire abstract derivation refer \cite{Madhu}],
\begin{equation}
DSI_x = {\left[1-\Big|
\frac{x-x_0}{x+x_0}\Big| \right]^{w_x}}\,,
\label{eq:esi}
\end{equation}
where $x$ is the concentration of the trace gases under observation, and $x_0$ is the standard for the respective trace gases as advised by Central Pollution Control Board (CPCB), India . \\ 
 
 The yearly Delhi Similarity Index is the geometric mean of the DSI value of the trace gases  (\ref{eq:esi}). 
 \begin{equation}
DSI ={[{DSI_{O_3}} \times {DSI_{SO_2}} \times {DSI_{CO}}]^{1/3}}
\end{equation}

Where, $DSI_{O_3}$, $DSI_{SO_2}$, and $DSI_{CO}$ are the Delhi Similarity Index values of ozone, sulfur-dioxide and carbon-monoxide, which ranges from 0 (dissimilar to Delhi) to 0.9-1 (similar to Delhi).
\label{sec:headings}

\section{Data Collection and Site Description}
\subsection{New Delhi}
The average altitude of Delhi is about 216m and the data for New Delhi is obtained from Central Pollution Control Board (CPCB)  \footnote{(\url{http://www.cpcb.gov.in/CAAQM/frmUserAvgReportCriteria.aspx})}, and this AQI data station is located at Shadipur ($28\degree 39'05.8"N, 77\degree 09'29.5"E$) India.
New Delhi is one of the most populated metropolitan region in India.
According to the 2011 census done by the Indian Government, the population  density of Delhi is about 11,000 people per square kilometer, 
The last census was taken 8 years ago, and given the exponential rise in population being observed in Delhi, the United Nations estimates that Delhi will become the most populated city by 2028.
The main reasons are the number of vehicles in Delhi has grown to a whooping 10 million
(Gude et al. 2006). The geopolitical disadvantage of New Delhi is; it is surrounded by industrial \& agricultural cities, which annually experiences a massive intake of trace gases as a direct effect of crop burning in neighbouring states.

\subsection{Bengaluru}
The altitude of Bengaluru is about 920 m and the data for Bengaluru is also obtained from CPCB website, the data station is located at BTM layout ($12\degree 54'47.5"N, 77\degree 36'33.3"E$) which is a residential area in Bengaluru. Unlike New Delhi, Bengaluru experiences a mild temperature throughout the year with almost no extreme variations; the annual mean temperature is around  25$^\circ$ C, April and May being the hottest.
Bengaluru is the target center of our study, which acts as the neutral point between New Delhi (major air pollution) and Jungfraujoch (minor air pollution). 

\subsection{Jungfraujoch}
Jungfraujoch is a station in the European Alps of Switzerland, it is located almost at the center of Europe, at an altitude of around 3460 m. It is a significant contributor to atmospheric data due to lack of pollutants at a good elevation and the nearby industrial areas located at lower altitudes. Data for Jungfraujoch was collected from the World Data Center for Greenhouse Gases [WDCGG] \footnote{(\url{https://ds.data.jma.go.jp/})} from the station monitored by Swiss Federal Laboratories for Materials Science and Technology (empa). 

Daily data collected, for all the three cities, from 2011 to 2014, the annual average is taken separately for all the trace gases and tabulated (Table: 2).


\section{Results}
The calculated weight exponent parameters for New Delhi data are tabulated in Table \ref{table:dsi_gas}. The weight exponents  calculations are done using  Eq \ref{eq:weight_exponent}, and the lower and upper limit of the trace gas data set is selected (here the minimum and maximum data recorded from 2011 - 2014). 

\begin{table}[h!]
\begin{center}
\caption{DSI Parameters' Table }\label{table:1}
\begin{tabular}{l*{4}{c}r}
\hline
Location & Gas & $x_0$ ($\mu gm^-3$) & Weight Exponents \\\hline
New Delhi & $O_3$	& 100  & 0.12\\ 
New Delhi & $SO_2$	& 50  & 0.10\\ 
New Delhi & $CO$	& 2000  & 0.081\\ 
\hline
\end{tabular}
\end{center}
\end{table}

Based on the exponents in Table \ref{table:1}, we calculate the Delhi similarity index for the the three trace gases using Equation \ref{eq:esi}  


\begin{table}[h!]
  \centering
  \caption{Delhi Similarity Index values of individual trace gases}
  \begin{tabular}{c c c c|c c c|c c c }
    \hline
    
\multirow{3}{1cm}{\textbf{Year}} &
\multicolumn{3}{c}{\textbf{New Delhi}} &
\multicolumn{3}{c}{\textbf{Bengaluru}} &
\multicolumn{3}{c}{\textbf{Jungfraujoch}}\\
    & \textbf{$CO$} & \textbf{$SO_2$} & \textbf{$O_3$} &
      \textbf{$CO$} & \textbf{$SO_2$} & \textbf{$O_3$} &
      \textbf{$CO$} & \textbf{$SO_2$} & \textbf{$O_3$}       
      \\\hline
    
    2011 & 0.983 & 0.901 & 0.922 & 0.885 & 0.815 & 0.946 & 0.085 & 0.561 & 0.998 \\ 
    2012 & 0.985 & 0.935 & 0.921 & 0.826 & 0.985 & 0.909 & 0.063 & 0.544 & 0.999 \\ 
    2013 & 0.961 & 0.912 & 0.909 & 0.893 & 0.972 & 0.654 & 0.065 & 0.546 & 0.998 \\ 
    2014 & 0.867 & 0.911 & 0.927 & 0.828 & 0.819 & 0.871 & 0.098 & 0.569 & 0.998  \\ \hline
  \end{tabular}
  \label{table:dsi_gas}
  
\end{table}

Table \ref{Tab:overall_dsi} shows the global Delhi Similarity Index estimated from the geometrical mean  of $CO$, $SO_2$, and $O_3$ for the three cities across 2011-14 data. 

\begin{table}[h!]
  \centering
  \caption{Global DSI values}
\begin{tabular}{l*{4}{c}r}
    \hline
Year & Delhi & Bengaluru & Jungfraujoch\\\hline
2011 & 0.934 & 0.880 & 0.691 \\ 
2012 & 0.947 & 0.904 & 0.689 \\
2013 & 0.927 & 0.828 & 0.689 \\
2014 & 0.901 & 0.839 & 0.697  \\ \hline 
  \end{tabular}
    \label{Tab:overall_dsi}
\end{table}


As evident from Table \ref{Tab:overall_dsi}, the DSI values for Delhi is maximum for all the years, whereas it is least for Jungfraujoch. The variation in Jungfraujoch is because of high $O_3$ content observed can be attributed to its high altitude.
The increase in the DSI value for Bengaluru is because of the increase in $SO_2$ concentration in 2013-14.\\

The concentration variations of trace gases from 2011 to 2014 data is graphically represented in figure 1. The large variations of data is observed in $CO$ and $O_3$, than $SO_2$. This could be because of large quantity of carbon emissions from vehicles, which in turn deplete the ozone layer.

\newpage
\begin{figure}[h!]
\centering        
\includegraphics[width=13cm,angle=0]{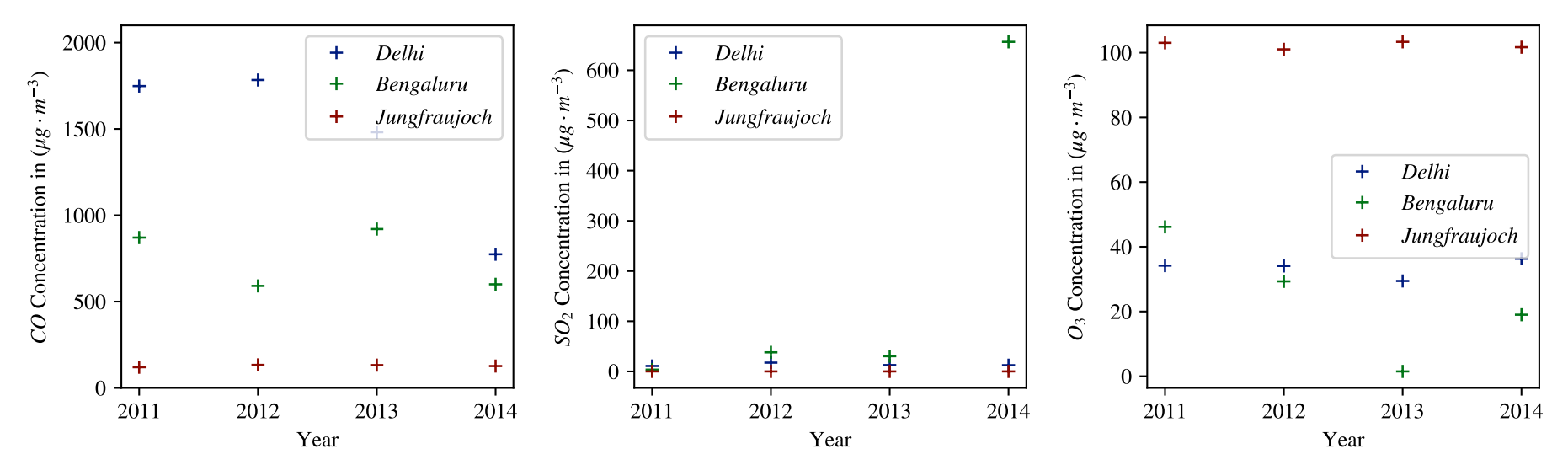} 
\caption{Concentration of trace gases $CO$, $SO_2$ and $O_3$ } 
\label{fig:1}
\end{figure}
The DSI results of Table\ref{Tab:overall_dsi} is graphically obtained as figure 2.

\begin{figure}[h!]
\centering        
\includegraphics[width=10cm,angle=0,scale=0.9]{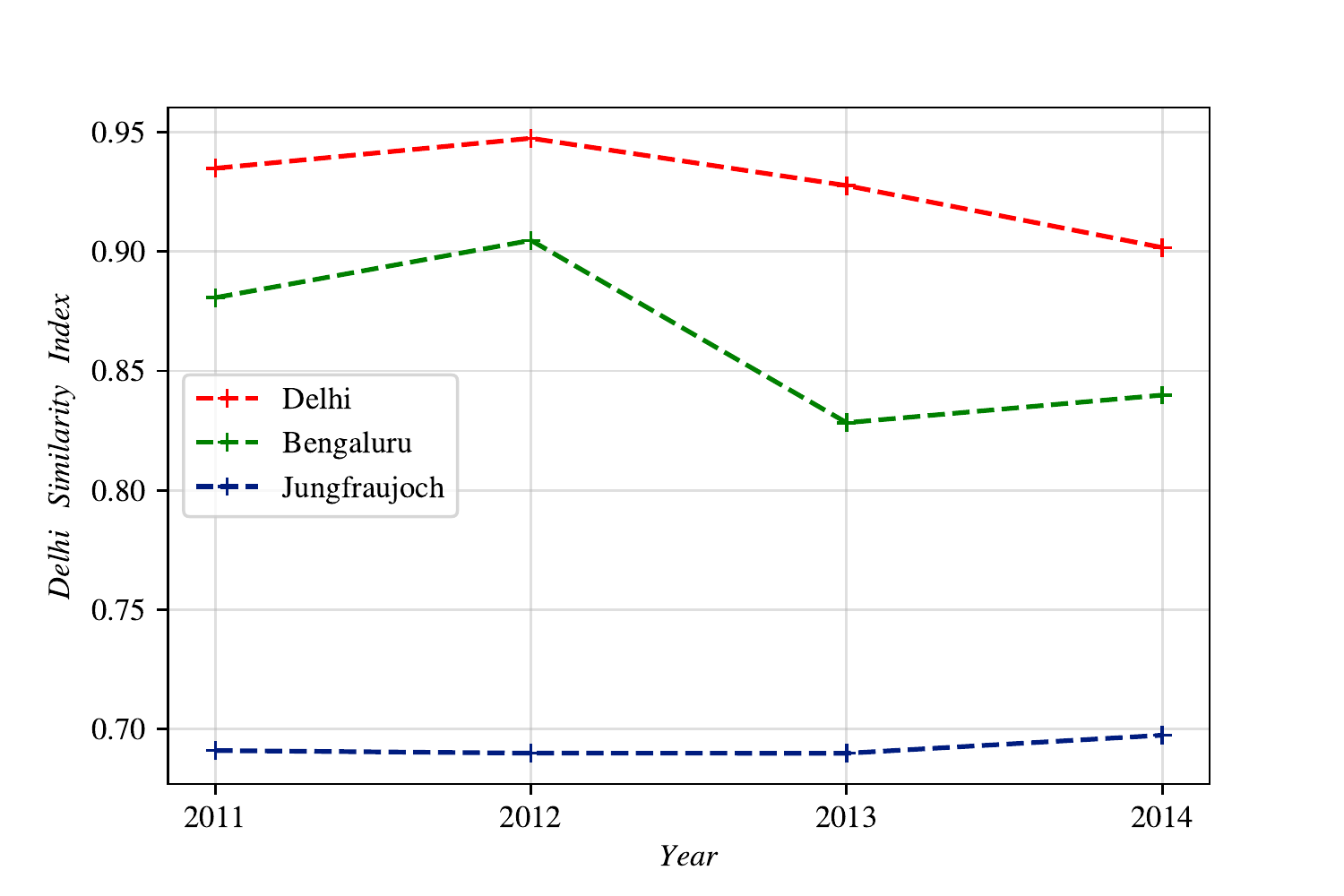} 
\caption{DSI values for Delhi, Bengaluru, and Jungfraujoch (2011-14)}
\label{fig:2}
\end{figure}

From figure 2 an increase is seen in Jungfraujoch and Bengaluru DSI values for the year 2013-14, while a downward trend is observed in Delhi DSI value. Nevertheless, the index value of Delhi remains prominent.

\subsection{$NO_2$ data deviation}
The boundary conditions of similarity index observed as a limit, that the data obtained from the CPCB database doesn't work on specific forms of data sets. The data set of $NO_2$ falls under the following condition: $x_{max} > x_{0}$ 

Where $x_{max}$ is the maximum value of x, and  $x_{0}$ is the reference value. The weight exponent calculated for this form of condition does not yield any result. Hence $NO_2$ data is not applied in the similarity index tool, but the tool is valid for the above-mentioned trace gases except $NO_2$.\\

\begin{figure}[h!]
\centering        
\includegraphics[width=10cm,angle=0]{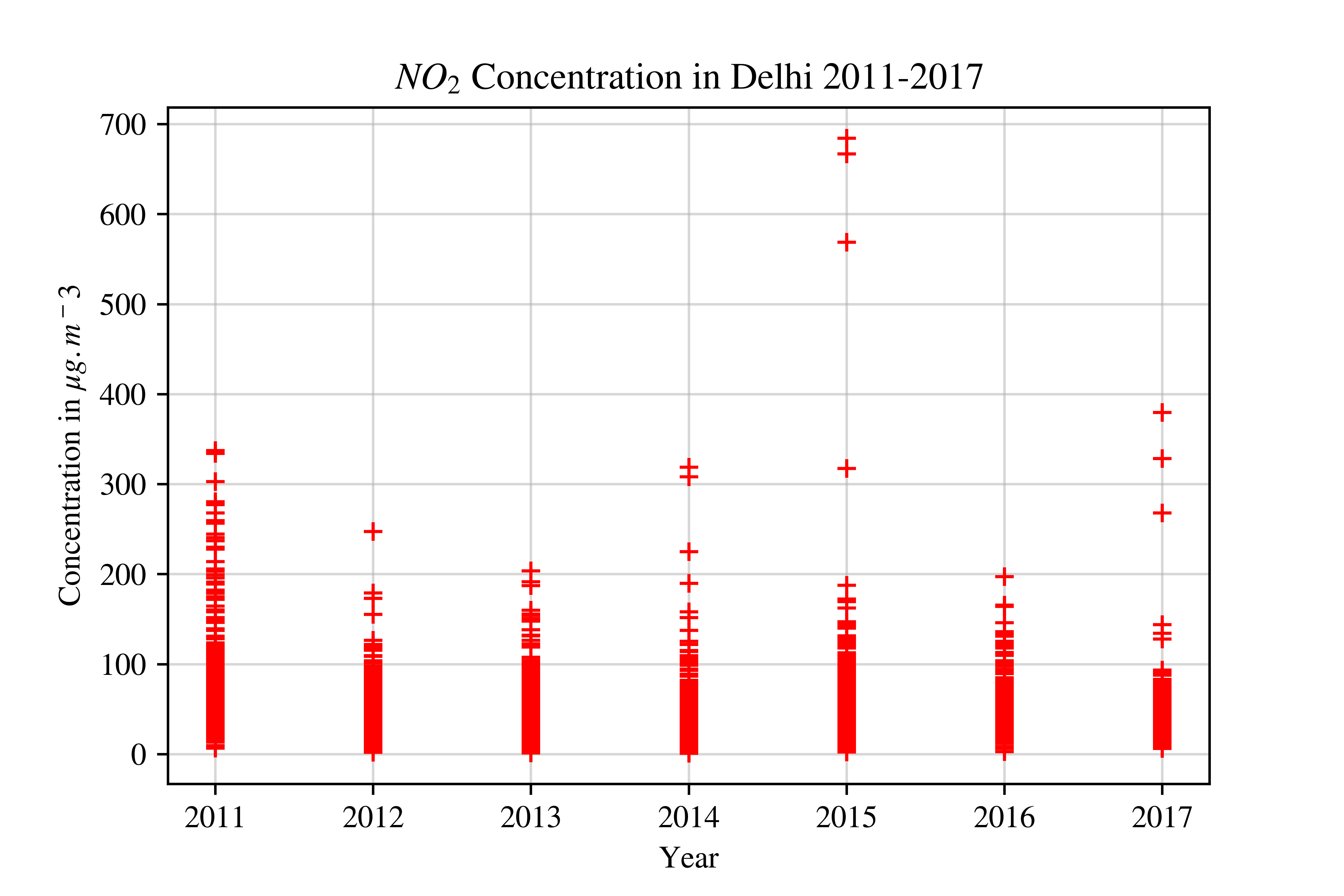} 
\caption{Daily concentration of $NO_2$ from 2011-2017.}
\label{fig:3}
\end{figure}

The Daily $NO_2$ concentration for Delhi is given in Figure \ref{fig:3}, a pattern is detected throughout 2011-17, with a high concentration of $NO_2$ occurring during November each year, which in turn can be linked to the burning of crops in the nearby cities.

The 2015 data corresponds to extremely high $NO_2$ values observed in first week of November in Delhi. Subsequently, the rest of the years follow the same pattern in November.

 
\section{PM 2.5 GIS analysis for Bengaluru and New Delhi}
Suspended particle matter in the air of size 2.5 is the basic full form of PM 2.5. The PM 2.5 data of the year 2017 from CPCB data is plotted using a Geographical Information system (GIS). The sites used for the analysis are Bengaluru and New Delhi, as data for Jungfraujoch, is not available. However, the ozone depletion is observed to be more significant in Jungfraujoch, than the PM 2.5 pollutant. GIS plots are used to study PM 2.5 effects and proliferation (as done in \cite{Jerrett}, \cite{Matejicek}).  

\subsection{Bengaluru}

\begin{figure}[h!]
\centering        
\includegraphics[scale=0.20]{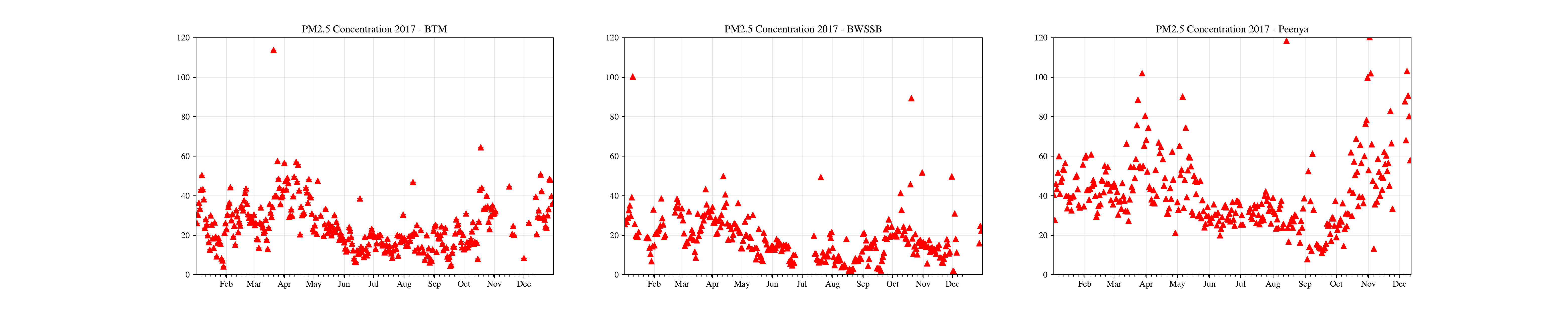} 
\caption{Daily PM 2.5 Value for three CPCB stations in Bengaluru}
\label{fig:cbcb_stations_bangalore}
\end{figure} 

Peenya, being an industrial district, harbours higher concentration of PM 2.5 when compared to residential areas (BTM, BWSSB). An increase in the relative content of PM 2.5 content in each district is observed in the month of April-May (Peak Summer months in Bengaluru). While a dip is observed in the month of June-July (onset of monsoon in Bengaluru). 

\begin{figure}[h!]
\centering        
\includegraphics[width=13cm,angle=0]{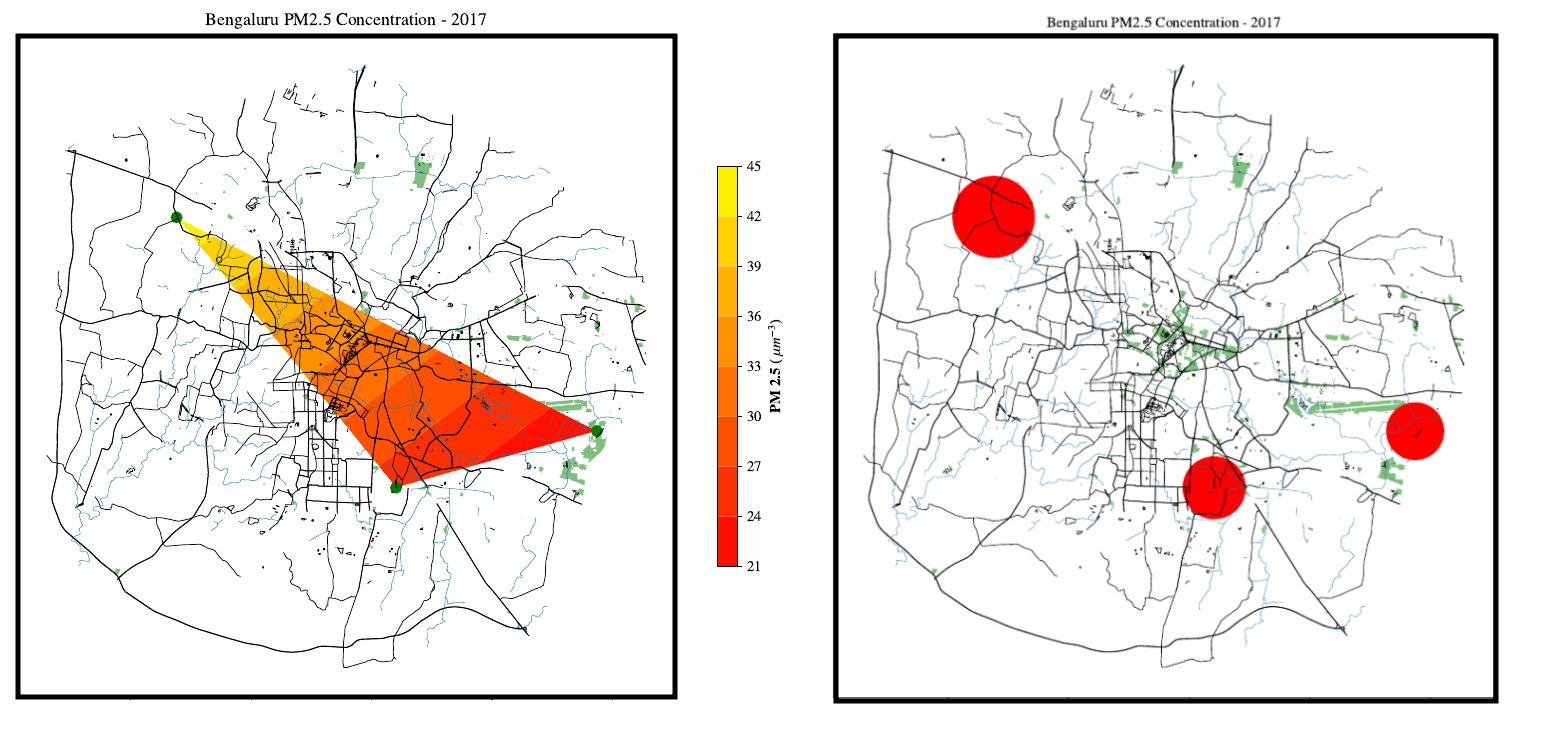} 
\caption{GIS interpolation of Bengaluru's annual PM 2.5 concentration from three CPCB stations across the cities (left). Relative scatter plot of Bengaluru for the same stations (right).}
\label{fig:5}
\end{figure}
The GIS plot (interpolation and scatter) for PM 2.5 content in Bengaluru, agrees with the above mentioned trends, like Peenya has higher concentration. The 3 stations are marked in green dots and the concentration intensity scale is shown in the right side of the GIS interpolation plot.

\subsection{New Delhi}
For the New Delhi PM 2.5 sample of 2017, the available data is taken from nine CPCB stations, (namely: ITO, Mandir Marg, Punjabi Marg, Dwarka, DTU, IHBAS, RK Puram, Shadipur, Sirifort) and the corresponding results are obtained as shown below.

\begin{figure}[h!]
\centering        
\includegraphics[width=8cm,angle=0]{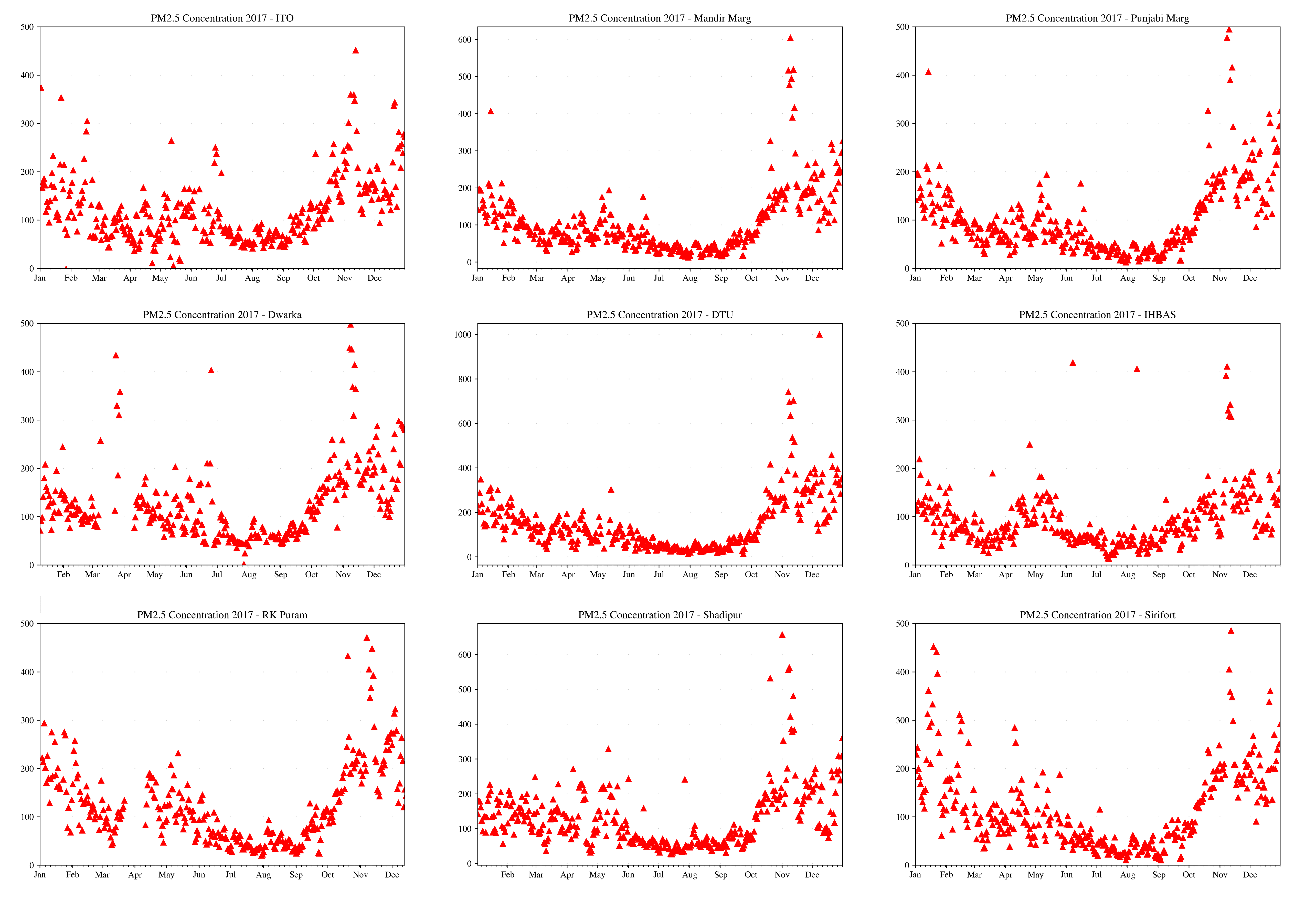} 
\caption{Daily PM 2.5 concentration plot for New Delhi, 2017 for nine CPCB stations}
\label{fig:delhi_cpcb_stations}
\end{figure}

The maximum PM 2.5 content is observed in the first week of November, which can be attributed to the burning of the crops in the neighbouring cities. The same pattern is observed in all the stations throughout the year. The PM 2.5 concentration is so high, that temperature does not seem to be the major factor in militating or mitigating the PM 2.5 content, as such seasonal variation was observed in Bengaluru (Figure \ref{fig:cbcb_stations_bangalore}). 


The GIS plot (interpolation and scatter plot) for New Delhi is given below in figure \ref{delhi_interpolation}.
\begin{figure}[h!]
\centering         
\includegraphics[width=13cm,angle=0]{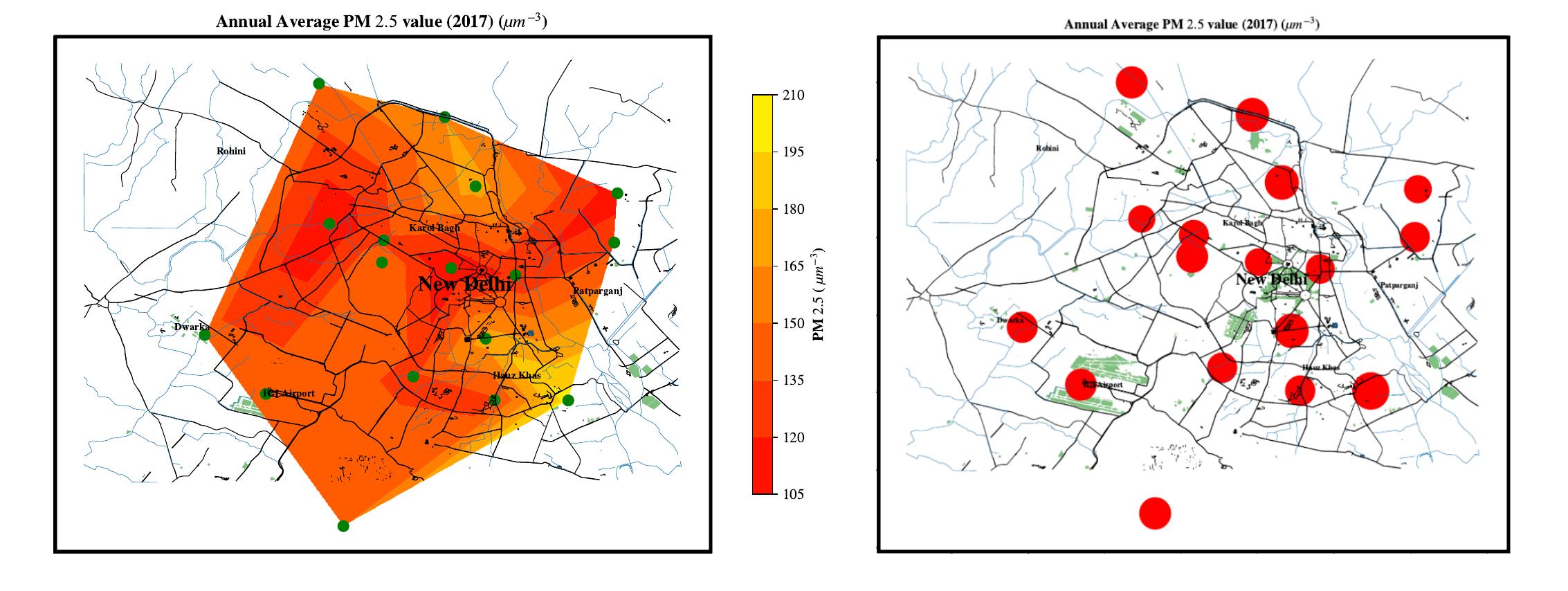} 
\caption{GIS interpolation of New Delhi's annual PM 2.5 concentration from around 17 CPCB stations across the city (left). Relative Scatter Plot of annual PM 2.5 average (right).}
\label{delhi_interpolation}
\end{figure}
The green dots in figure 7 indication CPCB stations and the variations of PM 2.5 intensity is high in  almost all chosen areas. The Bhuvan satellite vegetation images gives the size of built-up area (in red) of New Delhi and Bengaluru.

\begin{figure}[h!]
\centering         
\includegraphics[width=15cm,angle=0]{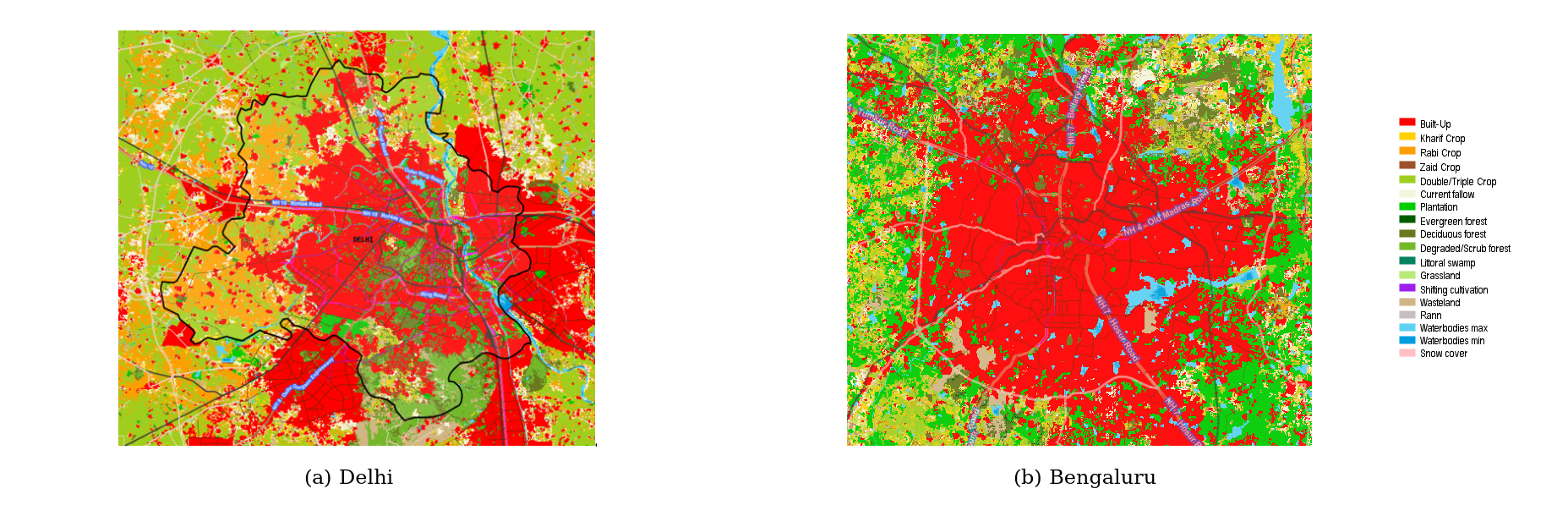} 
\caption{New Delhi and Bengaluru Vegetation with Bhuvan satellite view}
\label{fig:bhuvan_satellite}
\end{figure}

Even though the built up area of Bengaluru seems to be larger, the air pollution constraints seems to be follow the similar trend. The green and the blue, indicating the need of clean air and water has to be protected and improved. 
\newpage 
\section{Discussion and Conclusion}
As shown in Table \ref{table:dsi_gas},  different cities have different pollution levels for different gases, which is a result of human intervention, geographic position - altitude, and proximity from the coastline of the city. Therefore, a direct comparison of the concentration of trace gases is an inefficient measure of the level of pollution in a city. In other words, DSI is an index that can be used to compare the atmospheric deterioration (due to trace/greenhouse gases) for two or more cities. A succinct mathematical model, called Delhi Similarity Index (DSI), gives a reference value for comparison, interpolation and extrapolation. The cities that are chosen for the model are New Delhi, Bengaluru and Jungfraujoch. Wherein New Delhi is in severe pollution state, Jungfraujoch is in an ideal location in the Bernese Alps, and Bengaluru is taken as a neutral point between New Delhi and Jungfraujoch. Figure \ref{fig:1} and Figure \ref{fig:2} shows the variation of Concentration of the respective gases, and the overall DSI comparison of the three cities, respectively (from 2011-2014). As seen from Figure \ref{fig:2}, the year 2012 was worst, in terms of pollution, for the Indian cities of Bengaluru and New Delhi; after which both the cities started improving. The summary for all the 3 cities as given in the Table \ref{Tab:overall_dsi}. In turn implies that in 2011, the trace gases concentration Jungfraujoch was 69.1\% of New Delhi's on the same year, while that of Bengaluru's was 88\% of New Delhi's in the same year. Same can be stated about 2012, where a spike in New Delhi and Bengaluru DSI values are observed. similarly, the maximum value for Jungfraujoch was observed in 2014 (as seen in Figure \ref{fig:2}). Although Jungfraujoch's Carbon mono-oxide and Sulfur Dioxide content is of significantly lower amount when compared to New Delhi and Bengaluru, but the indexing gives 69\% similarity to New Delhi because of soaring surface ozone concentration in Jungfraujoch.
The variations the level of PM 2.5 due to seasonal changes (in Bengaluru) is an intriguing anomaly which requires further investigation, moreover New Delhi does not show such variation. Future work can be done with many more cities across a wide spectrum (coastal area, islands, tundra \& equatorial region, etc), with more trace / greenhouse gases data, with similarity index technique. A comprehensive comparison can then be made for the chosen cities.

\section*{Acknowledgments} 
We thank CPCB for there online database for data on Indian cities and  the World Data Centre for Greenhouse Gases (WDCGG)  database for Switzerland data. Also we would like to thank NRSC of ISRO for Bhuvan satellite data on vegetation.


\end{document}